\documentclass{ws-brlm}

\begin{document}

\markboth{J. Charvolin, J.-F. Sadoc} { Collagen and Phyllotaxis}


\title{A PHYLLOTACTIC APPROACH TO THE STRUCTURE OF COLLAGEN FIBRILS}

\author{JEAN CHARVOLIN and JEAN-FRAN\c COIS SADOC}

\address{Laboratoire de Physique des Solides,  Universit\'e Paris-Sud,
         CNRS UMR 8502,\\  F-91405 Orsay,cedex, France.\\
jean-francois.sadoc@u-psud.fr}

\maketitle

\begin{history}
\received{10 12 2010} \revised{31 03 2011}
To appear in Bio. Rev. Let.
\end{history}

\begin{abstract}
Collagen fibrils, cable-like assemblies of long biological molecules, the so-called triple helices, are dominant components of connective tissues. Their determinant morphological and functional roles motivated a large number of studies concerning their formation and structure. However, these two points are still open questions and, particularly, that of the lateral assembly of the triple helices which is certainly dense but not strictly that of a well-ordered molecular crystal. We examine here the geometrical template provided by the algorithm of phyllotaxis which gives  to each element of an assembly of points or parallel rods the most homogeneous and isotropic dense environment in a situation of cylindrical symmetry. The scattered intensity obtained from a phyllotactic distribution of triple helices in collagen fibrils presents features which could contribute to the scattering observed along the equatorial direction of their X ray patterns. Following this approach, the aggregation of triple helices in fibrils should be considered within the frame of soft condensed matter studies rather than that of molecular crystal studies.
%
%
\end{abstract}


\section{Introduction }

\begin{figure}[tbp]
\begin{center}
\includegraphics{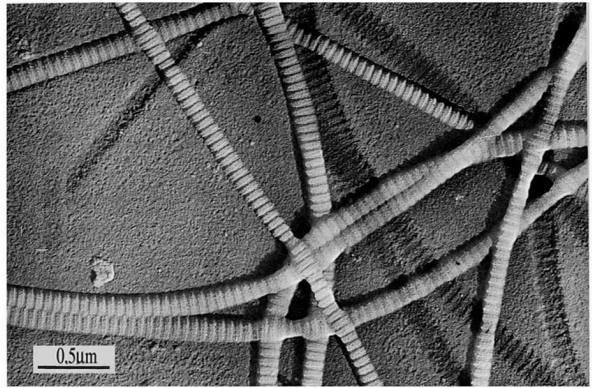}

\caption{Some fibrils of type I collagen observed with electron microscopy. Their diameter is of about 150 nm, the period of their longitudinal striations is of 67 nm (from J. Gross). These morphological features are rather constant whatever the conditions, in vivo as well as in vitro.}
\label{f1}
\end{center}

\end{figure}

\begin{figure}[tbp]
\begin{center}
\includegraphics{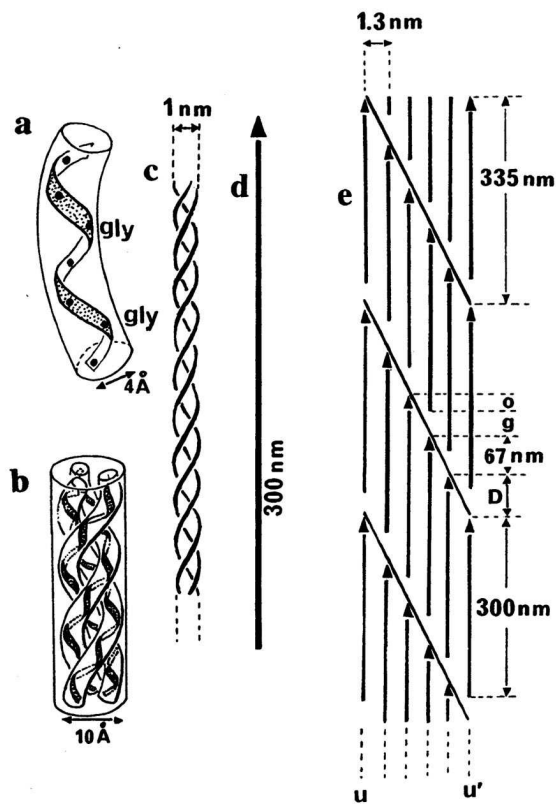}

\caption{Left-handed collagen molecules (a) assemble in right-handed triple helices (b,c) represented by rods (d) which in turn are proposed to be assembled with a regular shift (e) in order to create zones of gap `` g '' and overlap `` o ''  giving account of the striations. }
\label{f2}
\end{center}

\end{figure}

Collagen fibrils are the major constituent of connective tissues such as bone, cartilage, myofibrils, ligament, skin, cornea. Typical fibrils of type I collagen are shown in figure~\ref{f1}. Their properties and the rich polymorphism of their associations under different physico-chemical conditions allow the building of tissues adapted to the very diverse roles expected from them. Unfortunately, they can also underly the development of crippling deformities associated with rheumatic diseases and congenital defects in the tissues. The structure of collagen fibrils has therefore been the subject of many studies but is not yet totally solved. It is now commonly accepted that the formation of type I collagen fibrils proceeds along two steps as shown in figure~\ref{f2}: left-handed simple helices of collagen molecules build right handed triple helices which in turn build the fibrils\footnote{An intermediate step in which five triple helices would associate to build microfibrils was proposed earlier but is now questioned.}. If the first step is now well described and understood\cite{sadocrivier}, the second, the lateral organization of triple helices, is still subject of studies which can be summarized as follows.

\begin{figure}[tbp]
\begin{center}
\includegraphics{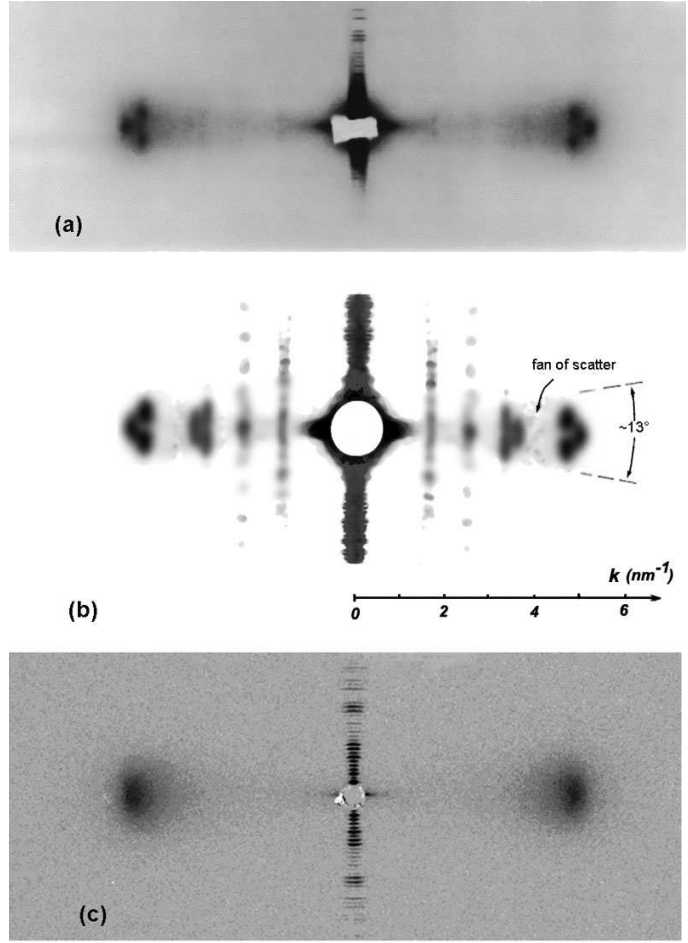}

\caption{Typical X ray pattern from  Hulmes  (a) accompanied by a drawing (b) strongly  emphasizing the observed scattering (for instance see also Wess and Orgel)  and another pattern from  Doucet   obtained under different conditions of preparation of the sample (c). The scattering along the equatorial direction is observed for wave vector $k\leq 5 $ nm$^{-1}$ only, a fan-like background is always present. The scattering along the meridional direction is due to the longitudinal striations. Here $k$ is defined by $k=4 \pi \sin \theta/\lambda$. }
\label{f3}
\end{center}

\end{figure}

The main information concerning the lateral organization of triples helices in fibrils is provided by X-ray scattering spectra whose intensities are directly related to the square of the Fourier transform of the organization. Typical X ray patterns are shown in figure~\ref{f3}: one from Hulmes\cite{hulmes81}, a drawing showing the results of background eliminations to put in light weak localized features as proposed by Wess and Orgel\cite{wess,orgel} and another pattern from  Doucet\cite{doucet}.

These examples  show that all the intensity is concentrated in the wave vector range $k\leq 5$nm$^{-1}$ only.
In all cases, an intense scattering is observed for $k\approx 5$nm$^{-1}$ which is to be related to the packing distance between triple helices as, if they were considered as straight parallel rods with a $1$ nm diameter packed at a distance $d=1.3$ nm along a hexagonal lattice, this should be the position of the first Bragg peak. In all cases also an important fan-like background scatter, including equatorial central scatterings out of the beam-stops, is observed for $k\leq 5$nm$^{-1}$ indicative of an important degree of disorder. However, weak localized spots are apparent on the first one\cite{hulmes81} which are not visible on the second\cite{doucet}. These spots reveals that an organization with some degree of lateral order on distances larger than the intermolecular one is possible, but their absence in the second case suggests that this organization can break down easily. These two extreme spectra most likely illustrate the sensitivity of the organization to the conditions of preparation and observation of the samples. They also show a subtle interplay between order and disorder  whose approach can be described in two steps schematically.

%
%
%
%

%
%
%
%

First, in order to obtain scattering in the wave vector range $k \leq5$nm$^{-1}$, periodic lattices with large parameters  respecting a dense packing of the triple helices close to that of a hexagonal lattice were proposed. For instance periodic deformations of a hexagonal lattice into a ``quasi hexagonal'' lattice with monoclinic symmetry\cite{hulmes79}  or a square-triangle lattice\cite{bouligand} \footnote{This square-triangle lattice is usually labeled $(3^2434)$ a notation pointing out the number of polygons around each of its vertices.}  inspired by the crystalline structures of model polypeptides\cite{berisio}  were proposed. Second, in order to give account of the broadening of the discernable spots and of the presence of a background, some type of disorder was introduced in the organization. This was done considering a quasicrystalline square-triangle tiling\cite{sasisekharan}  or paracrystalline fluctuations in a hexagonal lattice when the diffuse background is dominant\cite{doucet}.

%

It must be noted that in all cases except the last  the presence of a lattice should manifest itself as facets on the surface of the fibrils\footnote{Such facets are quite visible on toroidal aggregates resulting from DNA condensation, aggregates in which DNA is organized according to a hexagonal lattice\cite{hud,JChJFSdna,leforestier}.}
which are never observed whatever the diameter of the fibrils\cite{prockop}. Their circular section was indeed emphasized as a major morphological fact quite early\cite{ramachandran,woodhead}.
We propose here to consider that triple helices in fibrils adopt a lateral organization described by the algorithm of phyllotaxis, well known from botanists as a self-organized growth process and whose mathematical background is now well established\cite{coxeter,turing,jean}. This algorithm might be well adapted here insofar as it leads to organizations with cylindrical symmetry in which each element has the most homogeneous and isotropic environment,  in the same spirit as in ref.~\refcite{hulmes95}.  After having recalled the basis of phyllotaxis and made use of its algorithm as a template for overlap and gap regions of a fibril, we examine how the scattered intensities obtained from them can contribute to the analysis of the scattering data and to what extent phyllotaxis could provide a basis for the development of a structural model.

\section{Phyllotaxis}

\begin{figure}[tbp]
\begin{center}
\includegraphics{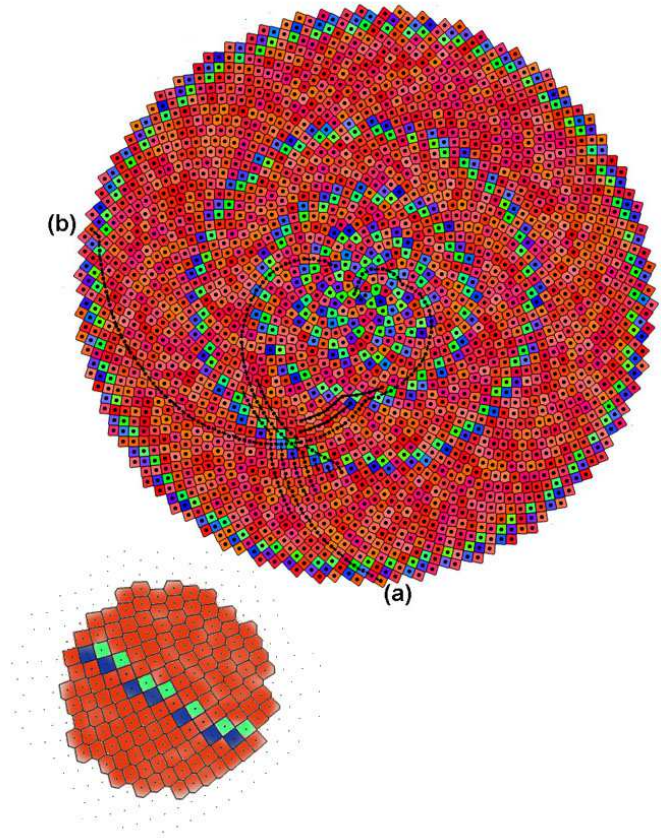}

\caption{A set of 2500 points organized according to the algorithm of phyllotaxis with the golden ratio. Each point is surrounded by its polygonal Voronoi cell whose number of sides corresponds to the number of first neighbors around the point. Blue cells are pentagons, red hexagons and green cells are heptagons. Those polygons are not necessarily regular and pentagons and heptagons are topological defects in the assembly of hexagons. Inserted on the left is a part of the figure showing in detail the evolution of Voronoi cells when crossing a ring of such defects. Representatives of two families of parastichies spirals are  drawn, (a) being right-handed and (b) left-handed by convention. Moving from the core towards the periphery, new parastichies are introduced when crossing a circle of defects.}
\label{f4}
\end{center}
\end{figure}

A phyllotactic organization of points indexed by $s$ is described by an algorithm such that the position of point $s$ is given by its polar coordinates
$r=a \sqrt{s}$ and  $\theta=2\pi \lambda  s$ that is $r=(a/ \sqrt{2\pi\lambda  })\sqrt{\theta}$  which is the equation of a Fermat spiral here called the generative spiral. As the circle of radius $r$ and area  $\pi r^2=\pi a^2 s$ contains $s$ points, the area per point has  the  value
 $\pi a^2$  (indeed it oscillates close to this value for small $s$ and then converges towards it). It has been shown\cite{ridley} that the the most homogeneous and isotropic environment, or the best packing efficiency in cylindrical symmetry, can be obtained  for $\lambda=1/\phi$, where $\phi$  is the irrational golden ratio\footnote{The two values $\lambda=1/\phi$ or $\lambda=1/\phi^2$ give two mirror images of the same points distribution as angles have to be considered modulo $2\pi$ and because $\phi^2= \phi+1$.}  $(1+ \sqrt{5})/2$. The phyllotactic organization of a set of 2500 points with their Voronoi cells is shown on figure~\ref{f4}.

%
%
%
%
The generative spiral is scarcely discernible on this figure, the radius vectors of two consecutive points being at an angle  $2\pi \phi $  modulo $2\pi$ or  $-137.5$ degrees, but the process generates quite conspicuous spirals, the so-called parastichies. There are three families of such spirals, each going through two opposite sides of a hexagonal Voronoi cell, whose representatives for two of them with opposite chiralities are drawn on the figure. The code chosen for the Voronoi cells exemplifies their remarkable degree of organization. Pentagons and heptagons are concentrated in narrow circular rings with constant width  which separate large rings of hexagons whose width increases progressively as one moves from the core to the periphery of the pattern. In the narrow rings pentagons and heptagons are associated in dipoles separated by hexagons whose shape is close to that of a square which would have two corners cut. From ring to ring, these dipoles are alternatively aligned along the spirals of one family of parastichies or the other. The counts of the numbers of dipoles and hexagons in each narrow ring and of the numbers of parastichies in each large ring, collected in table~1, put forward the mathematical substratum of the organization.
All the numbers listed in the table are indeed members of the Fibonacci series, $1,1,2,3,5,8,13,21,34,55,89,\ldots $, for which each number of rank $u$ is defined as $f_u=f_{u-1}+f_{u-2}$ from $f_0=0,f_1=1$. The number of sites in a narrow ring being that of its polygons, this number is also a Fibonacci number so that the radii of the narrow rings and the widths of the large rings are determined by the Fibonacci series. Moreover the distribution of dipoles on each ring is typical of that  of approximants of unidimensional quasicrystals\cite{rivier,rivier2}.

\begin{table}

\begin{center}

 \tbl{When going down the table, one moves from the core to the periphery of figure~\ref{f4}.  All numbers appearing in this table are successive Fibonacci numbers. This table stays limited to the beginning of the series. }
{\begin{tabular}{|c|c||c|c|}
\hline
\multicolumn{2}{|c||}{Narrow rings} &  \multicolumn{2}{c|}{Large rings} \\
\hline
{Dipoles of } &   & {Right handed } &  {Left handed } \\
{5 and 7-gons } & {hexagons} & { parastichies} &  {parastichies} \\
\hline
13&8& & \\
\hline
 & &13 &21 \\
\hline
21&13& & \\
\hline
 & &34&21\\
\hline
55&34& & \\
\hline
 & &34&55\\
\hline
55&34& & \\
\hline
 & &89&55 \\
\hline
89&55& & \\

\hline

\end{tabular}}
\end{center}
\end{table}

\begin{figure}[tbp]
\begin{center}
\includegraphics{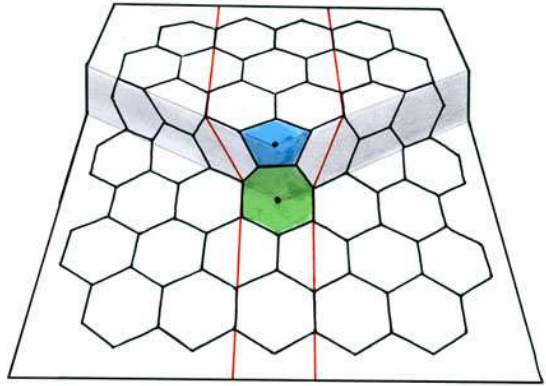}

\caption{One pentagon and one heptagon associated in a dipole create a dislocation in a bi-dimensional crystal of hexagons. The hexagons being kept regular the crystal  is deformed by a step. }
\label{f5}
\end{center}
\end{figure}

One observe also that when a narrow ring is crossed, moving from one large ring to the following, the number of parastichies increases for one family only by exactly the number of dipoles contained in the narrow ring. Thus, each dipole aligned along one family of parastichies generate one more parastichy in this family, an effect quite equivalent to that of a dislocation adding one more reticular plane in  part of a crystal and represented in figure~\ref{f5}. It was indeed proposed to extend here the language of metallurgy, speaking of the large rings as hexagonal grains and of the narrow rings separating them as grain boundaries\cite{rivier}. The role of those dipoles is essential to ensure the best packing efficiency as the points of the new parastichies they introduce enable the Voronoi cells to maintain their area and keep the most isotropic shape possible.

%
%
%
%
%

Thus, the local configuration is reproduced from one large ring to the next as it results from approximants in self-similar structures, this would not be the case if $\lambda$  was a rational number\cite{rivier,ridley}. Such an organization must be understood considering two kinds of disorders: a metric disorder corresponding to the fluctuations of distances between points with six first neighbors, their hexagonal cells being not regular everywhere, and a topological disorder associated to the introduction of pentagonal and heptagonal cells associated in dipoles. Those defects build concentric quasicrystalline rings with radii determined by the Fibonacci series so that an order is superposed onto the disordered lattice. This ordering within a disordered structure corresponds to the fact that, as the phyllotactic organization is defined by a single equation, it has, from the point of view of information theory, a low entropy\footnote{It is remarkable that this solution appears in botanical organizations not because a long evolutionary process has optimized the structures but because it is the only one possible with respect to sequential growth in a self organizing process under geometrical constraints.}.

%
\begin{figure}[tbp]
\begin{center}
\includegraphics{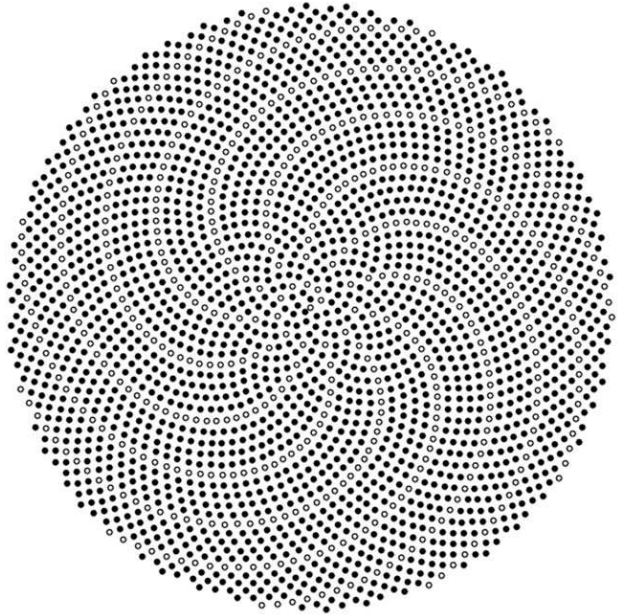}

\caption{Pattern with $N=2500$ points obtained when one point over five along the generative spiral represented by small circles  is suppressed in order to simulate vacancies in a gap region. }
\label{f6}
\end{center}
\end{figure}
%
%
%

\section{A phyllotactic template for collagen fibrils}

We consider a fibril as an assembly of parallel rods, the triple helices, organized following the algorithm of phyllotaxis in the normal section of the fibril, but this model must include the fact that the triple helices do not build continuous straight rods\footnote{A more realistic representation of triple helices in fibrils should include the presence of an eventual torsion as suggested by several observations, however it is expected to be weak and should manifest itself out of the equatorial trace. }. As shown in figure~\ref{f2}e,  the presence of striations along the fibrils led to propose that the intermolecular interactions of the triple helices put them in register so that each line is interrupted on 35 nm every 300 nm, those interruptions being regularly shifted by 35 nm from line to line. These interruptions correspond to vacancies within the assembly of triple helices. A fibril can therefore be modeled as an alternate stacking of small dense cylinders containing triple helix segments, overlap regions, and less dense small cylinders containing four triple helix segments out of five only, gap regions, all with a thickness of  35 nm.

%
\begin{figure}[tbp]
\begin{center}
\includegraphics{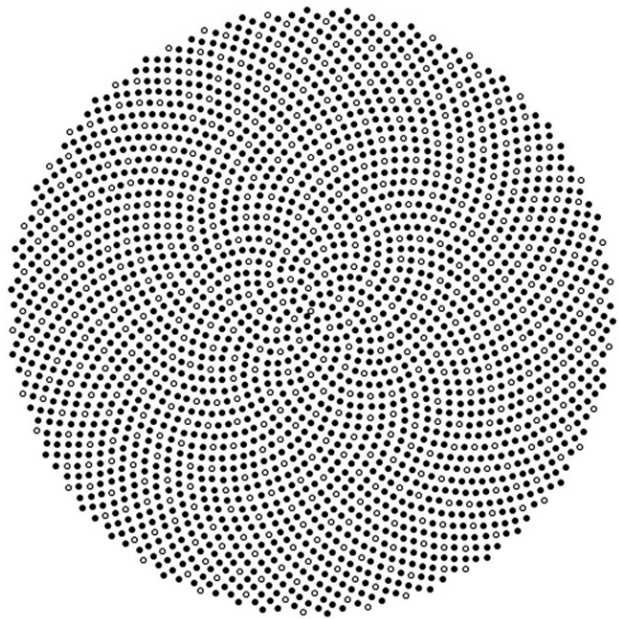}

\caption{Pattern with $N= 2500$ points obtained when one point over five within close groups of five points, represented by small circles, is suppressed in order to simulate vacancies in a gap region.  }
\label{f7}
\end{center}
\end{figure}
%

For the cylinders corresponding to overlap regions, the distribution of the triple helices in their normal plane is that of the phyllotactic pattern of figure~\ref{f4}, without modification. For the cylinders corresponding to the gap regions, we assume a similar pattern, but with missing points or vacancies owing to the fact that one point over five must be discarded. We examine here two possible distributions of vacancies in a gap region which can be considered as two extreme cases as far as the homogeneity of the distribution is concerned.

In the first case, the pattern is built eliminating one point over five along the generative spiral of the phyllotactic process. The result of this elimination is presented in figure~\ref{f6} in which eleven spirals of vacancies are separated by four spirals of points, a periodicity to be put in correspondence with the transverse repetition of
figure~\ref{f2}e.

In the second case, the pattern is built eliminating one point over five within groups of five points determined by a phyllotactic distribution of cells having an area five times larger than that of one point. This association of five triple helices is reminiscent of the microfibril model, but in which the microfibrils would have a phyllotactic organization. The result is presented in figure~\ref{f7} in which the vacancies are distributed in a much more homogeneous manner than in figure~\ref{f6}.

\section{Intensity scattered by a phyllotactic organization of triple helices}

As the triple helices are perpendicular to the phyllotactic patterns, modified or not by the presence of vacancies, the equatorial traces of the intensities scattered by the cylinders, gaps or overlaps, containing them are those of these patterns. The scattered intensity expected from a distribution of points is the square of the amplitude of its Fourier transform, but, as the points are representative of rods having a diameter of 1 nm, the points are to be changed into disks and the intensity decreases as the form factor of the disk. The results are shown on figure 8, for the overlap region without vacancy, figure 9, for a gap region whose vacancies are organized along spirals, and figure 10, for a gap region whose vacancies are distributed homogeneously.
%
%
%
%
\begin{figure}[tbp]
\begin{center}
\includegraphics{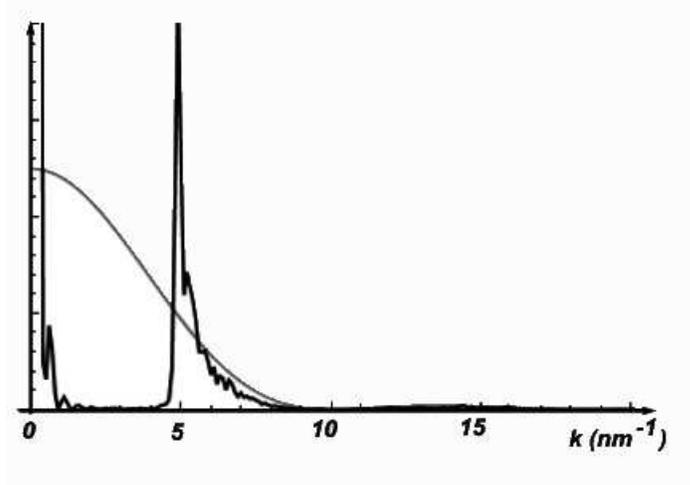}

\caption{  Radial average of the scattered intensity obtained from the phyllotactic pattern with $N=6200$ points, similar to that figure~\ref{f4}, overlap region, when its points are changed into disks having the diameter of the triple helices (1 nm), the form factor of the disk is the curve in grey.}
\label{f8}
\end{center}
\end{figure}
%
%

%
\begin{figure}[tbp]
\begin{center}
\includegraphics{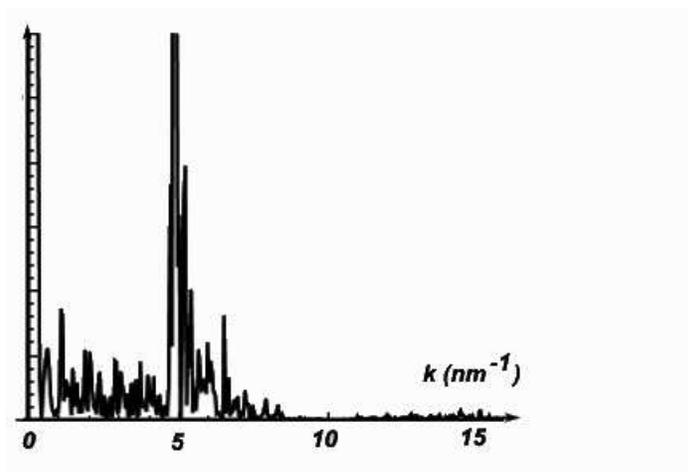}

\caption{ Radial average of the scattered intensity obtained from the pattern similar of that of figure~\ref{f6} (gap region with spiraling vacancies), with 6200 points instead of 2500. The arbitrary scale for intensities has been chosen to amplify small features at small $k$. }
\label{f9}
\end{center}
\end{figure}
%
%
\begin{figure}[tbp]
\begin{center}
\includegraphics{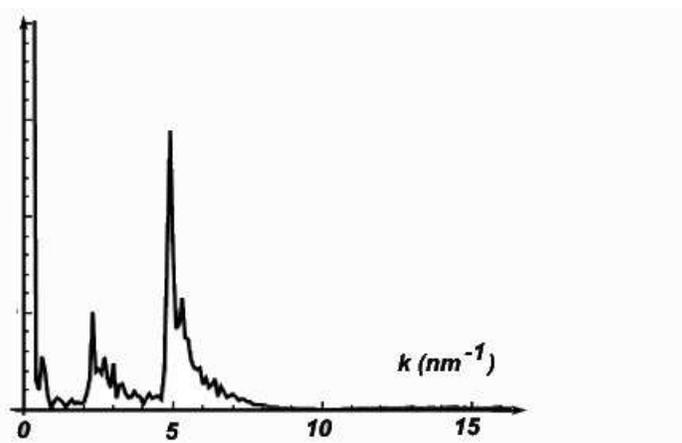}

\caption{ Radial average of the scattered intensity obtained from a pattern similar to that of figure~\ref{f7} (gap region with homogeneously distributed vacancies), with 6200 points instead of 2500.}
\label{f10}
\end{center}
\end{figure}
%
%
%
%
\begin{figure}[tbp]
\begin{center}
\includegraphics{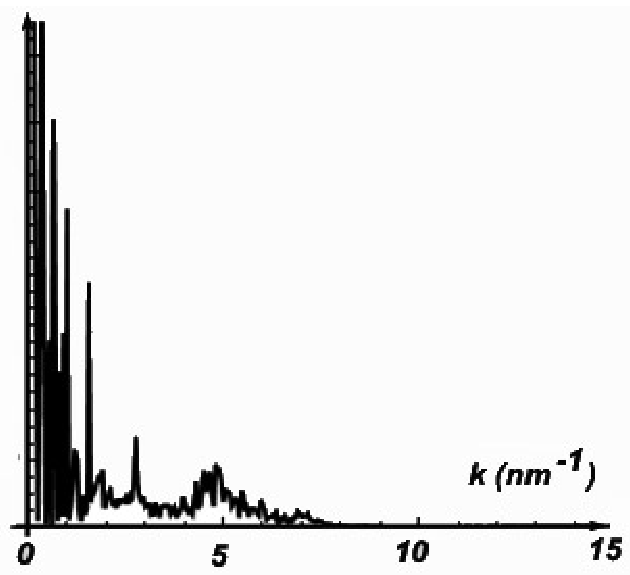}

\caption{Radial average of the intensity scattered by circles of dipoles only. The contribution of the long range correlations between these circles is quite visible in the range of lowest wave vectors only.  }
\label{f11}
\end{center}
\end{figure}
%
%
%

As for the spectra of figure~\ref{f3}, the scattering is concentrated in the wave vector range $k<5$ nm$^{-1}$, the absence of scattering for higher wave vectors is obviously due to the decrease of the form factor of the rod. Following an argument already used above, the quite intense and rather broad peak obtained for $k\approx5$ nm$^{-1}$ is indicative of a disordered local packing. For $k<5$ nm$^{-1}$, weaker structures are visible on the three traces which are to be related to the particular mode of organization built by the algorithm of phyllotaxis, i.e. the presence of rings of dipoles in the overlap as well as in the gap regions and that of vacancies in the gap region only. The first contribution can be identified considering figure 11 which shows the intensity scattered by a pattern in which all the points at the centers of hexagonal Voronoi cells outside the rings of dipoles are blanked out.

 One can infer from these figures that the total scattered intensity is the sum of several contributions. One around $k\approx5$ nm$^{-1}$ which corresponds to the local organization within the rings. A series of peaks for $k\leq 2.7$ nm$^{-1}$ which corresponds to the long range correlations between these rings\footnote{The scattering corresponding to the overall circular shape of the pattern should manifest itself, but at very small wave vectors, owing to the large value of its radius.}. Broader and more complex scatterings  for $2.3<k<5$ nm$^{-1}$ in figures 9 and 10  therefore corresponds to the organizations of vacancies in the gap region.
The contributions to the scattered intensity expected from the phyllotactic model of a fibril can therefore be summarized:\\
-	the presence of rings of dipoles in the overlap and gap regions for $k\leq2.7$ nm$^{-1}$, where their correlations lead to rather narrow scatterings whose separation decreases as $k\rightarrow 0$,\\
-	the presence of vacancies in the gap regions for $2.3\leq k \leq 5$ nm$^{-1}$, where their correlation lead to broader and more complex scatterings,\\
-	the intermolecular distance for $k \approx 5$ nm$^{-1}$, where the width of the peak reveals its local fluctuations.

\section{ Conclusions}

In a scattering pattern, the details which can be observed at a given value of the wave vector k have their origin in the density fluctuations within a mobile test  box of size proportional to $k^{-1}$. As, due to the form factor of the triple helix, the observations are limited to the wave vector range $0 <k <5$ nm$^{-1}$, the intensities observed for
$k <5$ nm$^{-1}$ corresponds to density fluctuations on a scale greatly larger than the intermolecular distance
and the phyllotactic model identify them as topological defects and vacancies.  The data presented in figure~\ref{f3}b could then be read in the following way:\\
-	the beamstop and the equatorial trace issued from it at small k mask or contain most of the contributions coming from the rings of dipoles and the two first spots observed can be put in correspondence, owing to their positions and narrow widths, with the two of these contributions obtained at larger k,\\
-	the third group of broader spots can be put in correspondence with the set of lines identified as coming from vacancies in the gap region, but with a preference for their spiral organization,\\
-	finally the last group of still broader and more intense spots reveals, as already said, the  disorder of the local molecular packing.

Those suggested correspondences between observed and calculated intensity curves should be more deeply investigated:\\
-	carrying on the development of the phyllotactic model considering noble numbers different from the golden ratio and the possibility of the melting of the distribution of dipoles\cite{rivier}, testing other distributions of vacancies in the gap region, taking into account the presence of a torsion and its influence on the off-equatorial scattering,\\
-	clarifying the influence of the conditions of preparation and observation on the structure, analyzing the central scattering which might contain information about eventual rings of dipoles and, finally, studying different types of collagen, particularly those without striations which might bring new information about the presence or distribution of vacancies.

Up to now, the only thing which can be said is that the phyllotactic approach proposes a reasonable structural model somewhere at midway between the two extreme cases represented by figures 3b and 3c. This approach makes use of a very simple algorithm, which might be thought too simple to give account of such a complex structural problem as far as the relation between order and disorder is concerned. However, the interest of this algorithm  is that, in the template so built, a topological order and a molecular disorder coexist intrinsically. This template belongs to the category of some models quoted above which discard true crystalline structures.  It is also supported by the fact that very different systems facing similar constraints, compactness and radial symmetry, adopt this solution spontaneously, as discussed below.

\section{Comment on the growth process}

The fact that the simple algorithm of phyllotaxis was mostly used to describe vegetal growth calls into question its implementation in the case of collagen as the growth process of the fibrils, which proceeds by addition of triple helices at their periphery, differs from the one at work in plants.

When this algorithm gives account of spiral organizations in flowers, the growth results from the addition of successive florets at the center of the flowers, the oldest are pushed aside by the new ones and organize themselves on the generative spiral at azimuthal angles of $137.5$ degrees from each other. This was illustrated studying the distribution of ferrofluid droplets falling down in a silicone oil in presence of a vertical inhomogeneous magnetic fields with cylindrical symmetry. The droplets polarized by the field push aside  those already fallen in the oil and the positions of the successive droplets in the liquid evolve collectively towards that of the phyllotactic distribution characterized by the golden ratio each time a new one falls\cite{douady}. Recently a  similar behavior has been observed for bubbles on a circular water surface\cite{yoshikawa}.

The phyllotactic organization of flowers is certainly the most notorious reference, but it was also proposed to call for such an organization in a quite different problem, that of the  the formation of B\'enard-Marangoni convection cells in a liquid layer contained in a cylindrical container and submitted to a vertical temperature gradient. As the gradient increases a transition takes place at which cells of uniform size determined by hydrodynamics appear in the liquid layer. Pentagonal and heptagonal cells, most often associated in dipoles, are observed among the hexagonal ones commonly formed in non cylindrical geometry. Their appearance was analyzed assuming the building of a phyllotactic pattern in order to respect the requirements of homogeneity and isotropy of the cells in cylindrical symmetry, whatever their distance from the center of the container\cite{rivier,rivier2}.

These examples stress upon the importance of cylindrical symmetry and of the necessity for the elements of the organization to move with respect to each other. We therefore propose here that the fibrils have a certain internal fluidity, for their molecules to reorganize each time a new one aggregates, and that their surface tension imposes them a circular section.  Fibrils should then be considered within the frame of soft condensed matter studies rather than that of molecular crystals studies. If this were so, whereas in the cases of flowers or droplets the growth is generated from the center and in that of the convection cells the organization builds itself at a collective transition, this growth of fibrils by addition at their surface could also provide a third example of the power of the algorithm of phyllotaxis in structural studies.

\section*{Acknowledgments}
We dedicate this article to the memory of our friend and colleague Yves Bouligand, died on the 21st of january 2011, whose studies of the morphological role of defects in biological materials has stimulated our advance in this field for many years.  We also thank Jean Doucet for fruitful discussions on scattering studies of fibrous proteins and for providing us with  figure~\ref{f3}c.


\end{document}